\documentclass[11pt]{article}
\usepackage{epsfig}
\usepackage{amscd}
\usepackage{amssymb}
\usepackage{amsfonts}
\usepackage{amsmath}
\usepackage{graphics,epsfig,psfrag}

\newcommand{\ssc}{\scriptscriptstyle}
\newcommand{\be}{\begin{equation}}
\newcommand{\ee}{\end{equation}}
\newcommand{\bea}{\begin{eqnarray}}
\newcommand{\eea}{\end{eqnarray}}
\def \HT{{\widetilde{\mathcal H}}}

\begin{document}

\begin{titlepage}

\baselineskip =15.5pt
\pagestyle{plain}
\setcounter{page}{0}

\begin{flushright}
%work in progress
\end{flushright}

\vfil

\begin{center}
{\huge {\bf Time and Geometric Quantization}}
\end{center}

\vfil

\begin{center}
{\large A. A. Abrikosov, Jr.}\\
\vspace {1mm}
ITEP, Bol. Cheremushkinskaya, 25;
        Moscow, 117259 Russia \\
        e-mail: {\it persik@vitep1.itep.ru}
\vspace {1mm}

\bigskip

{\large E. Gozzi and D. Mauro}\\
\vspace {1mm}
Department of Theoretical Physics, University of Trieste \\
Strada Costiera 11, Miramare-Grignano 34014, Trieste\\ and INFN,  
Trieste, Italy\\
e-mail: {\it gozzi@ts.infn.it} and {\it mauro@ts.infn.it}
\vspace {1mm}
\vspace{3mm}
\end{center}

\vfil

\begin{abstract}
\noindent In this paper we briefly review the functional version of the Koopman-von
Neumann operatorial approach to {\it classical} mechanics. We then show that 
its quantization can be achieved by freezing to zero two Grassmannian partners
of time. This method of quantization presents many similarities with the one
known as Geometric Quantization.
\end{abstract}
\vfil
\end{titlepage}
\newpage

\section{Introduction}	

The most difficult problem facing theoretical physics is the one 
of quantum gravity, i.e. the problem of putting together {\it gravity}
and {\it quantum mechanics} in a consistent way. By ``consistent way"
we mean a manner to get a quantum theory of gravity free of the ultraviolet
divergences which plague Einstein gravity when one tries to quantize it.

The strategy adopted over the last 25 years has been the one of attacking 
only the first horn of the problem that is {\it gravity}. This attack 
consisted in modifying classical gravity by adding new fields and interactions, 
even an infinite number of them, and obtaining models that go by the name 
of supergravities, strings, M-theory etc. Next one should quantize these models 
hoping to get finite or renormalizable results. This strategy does not seem to 
work even if it has produced a lot of nice physics and thousands of papers.

We wonder why people never thought of attacking the second horn of the problem 
which is {\it quantum mechanics} (QM). By ``attacking" we do not mean
that we have to modify QM, which is the best tested theory 
in the universe. ``Attacking" for us means giving a new look at QM especially
regarding its ``{\it geometrical}" aspects. After all if we want to put together 
QM with gravity, which is the queen of geometrical theories, we should better understand 
QM from a more geometrical point of view. What we mean is the following:
usually QM is formulated using tools like Hilbert space operators etc. while 
concepts related to space-time appear only at a secondary level or they do not appear at all.
This creates conceptual difficulties in applying QM to space-time theories like Gravity.

Actually ``{\it a space-time approach to quantum mechanics}" had already been explored by R.P. 
Feynman \cite{Feyn} in 1942-1948. His paper, which bears exactly the words above as title,
introduced for the first time the concept of path integral or functional integration into QM. 

In this paper we will continue on this line by introducing path integrals even at the {\it classical}
mechanics level and showing that quantization, that is the transition to the Feynman's {\it quantum}
path integral, can be achieved by freezing to zero two Grassmannian partners of time. We feel 
that this analysis of quantization is the one which brings concepts related to space-time
(or its partners) in closer contact with QM.

\section{Koopman-von Neumann Operatorial Approach to Classical Mechanics and Its Path Integral 
Version}

Quantum mechanics, at least in the formulation known in the early Thirties, was a theory
made of operators and Hilbert spaces while classical mechanics (CM) was a theory of 
commuting $c$-functions and phase spaces. Koopman and von Neumann \cite{koop} (KvN), 
triggered most probably by those differences above between QM and CM, proposed an 
operatorial formulation of CM. This
is basically an extension of the work of Liouville who 
had found the equation of motion for the probability distributions $\rho(q,p)$
on phase space:
%%%
\begin{equation}
\displaystyle i\frac{\partial}{\partial t} \rho(q,p)=\hat{L}\rho(q,p) \label{2-1}
\end{equation}
%%%
where $\hat{L}$ is the Liouvillian defined as
${\hat L}=-i\partial_pH\partial_q+i\partial_qH\partial_p$.
Koopman and von Neumann instead of using the space of probability distributions $\rho(q,p)$,
which are $L^{\scriptscriptstyle 1}$-functions, {\it postulated} the same equation (\ref{2-1})
for a set of functions $\psi(q,p)$ defined on phase space:
%%%
\begin{equation}
\displaystyle i\frac{\partial}{\partial t}\psi(q,p)=\hat{L}\psi(q,p). \label{2-2}
\end{equation}
%%%
These functions $\psi(q,p)$ are square integrable ($L^{\scriptscriptstyle 2}$-functions) and they
make up a Hilbert space.
Moreover KvN postulated that the $\rho$ 
could be built out of the $\psi$ as
%%%
\begin{equation}
\rho(q,p)=|\psi(q,p)|^2. \label{2-3}
\end{equation}
%%%
The first thing to notice is that, as $\hat{L}$ is first order in the derivatives, one can 
obtain Eq. (\ref{2-1}) as a consequence of (\ref{2-2}) and (\ref{2-3}). Note that 
this would not happen if instead of the Liouvillian $\hat{L}$ we would have the Schr\"odinger operator
$\hat{H}$ because this one is second order in the derivatives.

The one briefly exposed above is the KvN operatorial approach to CM \cite{nostri}. 
This formalism can be generalized to include the evolution of higher forms,
that means of objects like 
%%%
\begin{equation}
\psi(\varphi^a, d\varphi^a)=\psi_{\scriptscriptstyle 0}(\varphi)+\psi_a(\varphi)d\varphi^a
+\psi_{ab}(\varphi) d\varphi^a\wedge d\varphi^b+\ldots \label{7-0}
\end{equation}
%%%
where $\varphi^a=(q^{\scriptscriptstyle 1}\cdots q^n,p^{\scriptscriptstyle 1}\cdots p^n)$
with $a=1, \ldots, 2n$. 
The operator which makes the evolution of the $\psi$ above is a generalization 
of the Liouvillian and it is known as the Lie derivative of the Hamiltonian flow \cite{marsd}.
Having now an operatorial version for CM one could ask if, like for any operatorial formalism, 
there is a correspondent path integral version. The answer is yes \cite{enniocl}; 
basically it is a path integral which gives weight one to the classical path 
and zero to all the others. So the classical probability {\it amplitude}\,\footnote{We can call
$K$ a transition {\it amplitude} because it gives the kernel of evolution
of the ``wave functions" $\psi$
of the Hilbert space introduced by KvN. In order to distinguish these wave functions from the quantum
mechanical ones, we have called them ``Koopman-von Neumann Waves" \cite{nostri}.}
$K(\varphi_{\scriptscriptstyle f};
t_{\scriptscriptstyle f}|\varphi_{\scriptscriptstyle i};t_{\scriptscriptstyle i})$ 
of going from an initial phase space configuration $\varphi_{\scriptscriptstyle i}$
at the initial time $t_{\scriptscriptstyle i}$ to a final one 
$\varphi_{\scriptscriptstyle f}$ at time $t_{\scriptscriptstyle f}$ is given by
%%%
\begin{equation}
K(\varphi_{\scriptscriptstyle f};t_{\scriptscriptstyle f}|\varphi_{\scriptscriptstyle i};
t_{\scriptscriptstyle i}\bigr)
=\delta\bigl[\varphi_{\scriptscriptstyle f}-{\widetilde{\phi}}_{cl}(t_{\scriptscriptstyle f};\,
\varphi_{\scriptscriptstyle i}, t_{\scriptscriptstyle i})\bigr] \label{7-1}
\end{equation}
%%%
where $\widetilde{\phi}_{cl}(t;\, \varphi_{\scriptscriptstyle i}, t_{\scriptscriptstyle i})$ 
is the classical
trajectory starting at time $t_{\scriptscriptstyle i}$ from the phase space point
$\varphi_{\scriptscriptstyle i}$.
These trajectories are solutions of the Hamilton equations of motion:
$\displaystyle \dot{\varphi}^{a}=\omega^{ab}\frac{\partial H}{\partial\varphi^{b}}$ with
$\omega^{ab}$ the standard symplectic matrix and $H$ the Hamiltonian of the system.  Slicing the time
interval $t_{\scriptscriptstyle f}-t_{\scriptscriptstyle i}$ into $n$ pieces and doing some
manipulations on the Dirac delta appearing on the RHS of (\ref{7-1}), we can
rewrite the transition amplitude $K$ as a path integral \cite{enniocl}:
%%%
\begin{equation}
K\bigl(\varphi_{\scriptscriptstyle f};t_{\scriptscriptstyle f}
|\varphi_{\scriptscriptstyle i};t_{\scriptscriptstyle i}\bigr) =
\int{\cal D}{\mu}\, \exp \biggl[
i\int_{t_{\scriptscriptstyle i}}^{t_{\scriptscriptstyle f}}dt\,{\widetilde{\cal
L}}\biggr]. \label{8-1}
\end{equation}
%%%
The functional integration measure is 
${\cal D}\mu\equiv {\cal D}^{\prime\prime} \varphi^{a} {\cal D}
\lambda_{a} {\cal D}c^{a} {\cal D}{\bar c}_{a}$ where
${\cal D}^{\prime\prime}$ indicates that the integrations over $\varphi_{\ssc i}$ and
$\varphi_{\ssc f}$ are not done. The $6n$
auxiliary variables $\lambda_{a}$,\, $c^{a}, \,{\bar c}_{a}$ (of which
$c^a$ and ${\bar c}_a$ are Grassmannian) make their appearance 
via the transformations performed on the Dirac deltas \cite{enniocl}. 
The Lagrangian ${\widetilde{\cal L}}$ in (\ref{8-1}) and its
associated Hamiltonian are:
\begin{equation}
\begin{array}{l}
{\widetilde{\cal L}}=\lambda_{a}[{\dot\varphi}^{a}-\omega^{ab}
\partial_{b}H]+
i{\bar c}_{a}[\delta^{a}_{b}\partial_{t}-\omega^{ac}\partial_{c}\partial_{b}H]
c^{b}\smallskip \\
{\HT}=\lambda_a\omega^{ab}\partial_bH+i\bar{c}_a\omega^{ac}
(\partial_c\partial_bH)c^{b}. \label{8-2}
\end{array}
\end{equation}
%%%
All the auxiliary variables $(\lambda_a,c^a,\bar{c}_a)$ have a very nice geometrical
interpretation \cite{enniocl}, for example the $c^a$ can be identified with the basis 
of the differential forms 
$d\varphi^a$ appearing in (\ref{7-0}) and the Hamiltonian $\widetilde{\cal H}$
of (\ref{8-2}) with the Lie derivative of the Hamiltonian flow. The connection 
of (\ref{8-1}) with the KvN operatorial formalism can easily be established \cite{enniocl}. 
For example from the path integral (\ref{8-1}) one can derive
that the only graded commutators different from zero are:
%%%
\begin{equation}
[\hat{\varphi}^a,\hat{\lambda}_b]=i\delta_b^a,\qquad\qquad [\hat{c}^a,\hat{\bar{c}}_b]=
\delta_b^a. \label{comm}
\end{equation}
%%%
If we consider the representation in which $\hat{\varphi}^a$ is an operator of multiplication 
and $\hat{\lambda}_a=-i\partial_a$, then the first
term of the Hamiltonian $\widetilde{\cal H}$ of (\ref{8-2}) becomes
$\widetilde{\cal H}=-i\omega^{ab}\partial_bH\partial_a$ which is just the Liouvillian $\hat{L}$
appearing in the equation of evolution of the KvN waves (\ref{2-2}). This confirms 
that the path integral (\ref{8-1}) is the functional counterpart of the KvN theory.

\section{Grassmannian Partners of Time and Geometric Quantization}

The question we would like to address now is: ``{\it Having CM formulated via path integrals
how do we go to the quantum theory? In other words, how do we pass from the path integral
(\ref{8-1}) of CM (let us call it CPI for Classical Path Integral) to the Feynman's
quantum path integral (QPI):
%%%
\begin{equation}
\displaystyle 
\langle q_{\scriptscriptstyle f};t_{\scriptscriptstyle f}|q_{\scriptscriptstyle i};
t_{\scriptscriptstyle i}\rangle=
\int {\cal D}^{\prime\prime}q{\cal D}p \;\exp\biggl[\frac{i}{\hbar}S[\varphi]\biggr] \label{13}
\end{equation}
%%%
where $S[\varphi]=\int dt \, L(\varphi)=\int dt \,(p\dot{q}-H)$?}".
Let us first try to rewrite the weight of the path integral 
(\ref{8-1}) in a better form. To do that we will assemble the $8n$
variables $(\varphi^a,c^a,\bar{c}_a,\lambda_a)$ as follows. First we introduce
two Grassmannian partners $\theta,\bar{\theta}$ of the standard time $t$ appearing in (\ref{8-1})
and next we define the following functions of $(t,\theta,\bar{\theta})$:
%%%
\begin{equation}
\Phi^a(t,\theta,\bar{\theta})\equiv \varphi^a(t)+\theta c^a(t)+\bar{\theta}\omega^{ab}
\bar{c}_b(t)+i\bar{\theta}\theta\omega^{ab}\lambda_b(t).
\end{equation}
%%%
$\Phi^a$ is known in the literature as {\it superfield} \cite{salam} while the 
extension $(t,\theta,\bar{\theta})$ of the $t$-space is called {\it superspace}. Somehow
$\Phi^a$ puts together in the same ``{\it multiplet}" 
all the $8n$ variables of the CPI. 
The possibility of introducing this superfield in our formalism is
related to the fact that the Lagrangian $\widetilde{\cal L}$ of (\ref{8-2}) presents various hidden
BRS-like and susy-like symmetries \cite{enniocl}. 
These invariances are the representation on the $(\varphi,
c,\bar{c},\lambda)$-space of some symmetries at the superspace level which mix among 
themselves the $(t,\theta,
\bar{\theta})$-variables. The superfield $\Phi^a$ has been a very powerful tool in supersymmetric 
theories \cite{salam} especially at the perturbative level. In our formalism instead it helps in bringing 
to light the interplay between the CPI and the QPI. For example if we replace the variables 
$\varphi$ with the superfields $\Phi$ into the original Hamiltonian $H(\varphi)$ 
and expand it in $\theta,\bar{\theta}$ we get:
%%%
\begin{equation}
H(\Phi)=H(\varphi)+
\theta\frac{\partial H}{\partial\varphi^{a}}c^{a}+
\bar{\theta}\frac{\partial H}{\partial\varphi^{a}}\omega^{ab}{\bar c}_{b}+
i\theta\bar{\theta}\widetilde{\cal H}.
\end{equation}
Doing the same with the action $S[\varphi]$ appearing in the weight of the QPI
(\ref{13}), we get:
%%%
\begin{equation}
\label{11-1}
S[\Phi]=S[\varphi]+\theta {\cal T}+\bar{\theta}{\cal
V}+i \theta\bar{\theta}\, [{\widetilde {\cal S}}-(s.t.)]. 
\end{equation}
%%%
$\widetilde{S}$ is $\int dt\,\widetilde{\cal L}$ with $\widetilde{\cal L}$ given by (\ref{8-2}),
${\cal V}$ and ${\cal T}$
are functionals of $\varphi,c,\bar{c},\lambda$ of which it is not important to know 
the explicit form, while 
$(s.t.)$ is the following surface term: 
%%%
\begin{equation}
\displaystyle (s.t.)=(\lambda_{p}p
+i{\bar c}_{p}c^{p})\bigl|^{t_{\ssc f}}_{t_{\ssc i}}. \label{11-2}
\end{equation}
%%%
The important point to notice is that in the multiplet on the RHS of (\ref{11-1}) the first term 
$S[\varphi]$ is the weight entering the QPI while the last one, modulo
the $s.t.$, is the one entering the CPI. The occurrence, in the same supermultiplet, of both the
QM and CM actions cannot be an accident and must have some deeper meaning. 

Before proceeding we need to go back to (\ref{8-1}). 
That one was the kernel of transition between points
in $\varphi$-space but one could build also the kernel of transition between points in 
$(\varphi,c)$-space. The reason is that, at the operatorial level, $\hat{\varphi}$ and $\hat{c}$
commute and so they can be diagonalized simultaneously 
%%%
\begin{equation}
\hat{\varphi}|\varphi,c\rangle=\varphi|\varphi,c\rangle, \qquad\qquad 
\hat{c}|\varphi,c\rangle=c|\varphi,c\rangle.
\end{equation}
%%%
The transition amplitude $\langle\varphi_{\ssc f},c_{\ssc f};
t_{\ssc f}\vert\varphi_{\ssc i},c_{\ssc i};t_{\ssc i}\rangle$ will then have 
the following path integral expression:
\begin{equation}
\label{12-1}
\langle\varphi_{\ssc f},c_{\ssc f};
t_{\ssc f}\vert\varphi_{\ssc i},c_{\ssc i};t_{\ssc i}\rangle=
\int{\cal D}^{\prime\prime}\varphi
{\cal D}^{\prime\prime}c{\cal D}\lambda{\cal D}\bar{c}\; e^{i\widetilde{S}}.
\end{equation}
%%%
Next let us notice from (\ref{11-1}) that 
%%%
\begin{equation}
\widetilde{S}=\int \, id\theta d\bar{\theta}S[\Phi] +(s.t.).
\end{equation}
%%%
So (\ref{12-1}) can be rewritten as
%%%
\begin{equation}
\label{12-2}
\langle\varphi_{\ssc f},c_{\ssc f};
t_{\ssc f}\vert\varphi_{\ssc i},c_{\ssc i};t_{\ssc i}\rangle=
\int{\cal D}^{\prime\prime}\varphi
{\cal D}^{\prime\prime}c{\cal D}\lambda{\cal D}\bar{c}\,
\exp \biggl[i\int \,id\theta d\bar{\theta}S[\Phi] +(s.t.)\biggr].
\end{equation}
%%%
We can get rid of the surface term on the RHS of (\ref{12-2}) by performing 
a proper Fourier transform of both sides of (\ref{12-2}). The result of this 
Fourier transform \cite{wip} is that the $\varphi^a=(q,p)$-variables on the LHS
of (\ref{12-2}) are replaced by the $(q,\lambda_p)$-ones while the variables
$(c^q,\bar{c}_p)$ take the place of $c^a=(c^q,c^p)$. 
The index ``$q$" stands for the first $n$ indices ``$a$"
in $(\varphi^a,c^a,\lambda_a,\bar{c}_a)$ and ``$p$" for the second ones. 
After the Fourier transform we obtain:
%%%
\begin{equation}
\displaystyle 
\langle \Phi^q_{\scriptscriptstyle f};t_{\scriptscriptstyle f}|\Phi^q_{\scriptscriptstyle i};
t_{\scriptscriptstyle i}\rangle =\int {\cal D}^{\prime\prime}\Phi^q {\cal D}\Phi^p \,
\exp \biggl[i\int \,idtd\theta d\bar{\theta}L[\Phi]\biggr] \label{12-3}
\end{equation}
%%%
where ${\cal D}^{\prime\prime}\Phi^q={\cal D}^{\prime\prime} q{\cal D}^{\prime\prime}\lambda_p
{\cal D}^{\prime\prime} c^q{\cal D}^{\prime\prime}\bar{c}_p$,
${\cal D}\Phi^p={\cal D}p{\cal D}\lambda_q
{\cal D}c^p{\cal D}\bar{c}_q$
and $|\Phi^q;t_{\scriptscriptstyle i}\rangle$ stands for $|q,\lambda_p,c^q,\bar{c}_p;
t_{\scriptscriptstyle i}\rangle$ because both 
of them are eigenstates of the operator $\hat{\Phi}^q$ with eigenvalue $\Phi^q$. 
Remember also that $\Phi^q=q+\theta c^q+\bar{\theta}\bar{c}_p
+i\bar{\theta}\theta\lambda_p$ and so $\displaystyle \lim_{\theta,\bar{\theta}\to 0} \Phi^q=q$.
At this point we cannot avoid noticing the striking similarity between the relation (\ref{12-3})
valid for CM and the quantum one which is given by:
%%%
\begin{equation}
\displaystyle 
\langle q_{\scriptscriptstyle f};t_{\scriptscriptstyle f}|q_{\scriptscriptstyle i};
t_{\scriptscriptstyle i}\rangle=
\int {\cal D}^{\prime\prime}q{\cal D}p \;\exp\biggl[\frac{i}{\hbar}\int dt\,L[\varphi]\biggr]. \label{13-1}
\end{equation}
%%%
One goes from (\ref{13-1}) to (\ref{12-3}) by replacing $q\to \Phi^q$, $p\to \Phi^p$
and extending the integration measure from $\int dt$ to the whole superspace 
$i\hbar\int dt d\theta d\bar{\theta}$. The factor $i\hbar$ is crucial in order
to have a measure of integration real and with the correct physical dimensions:
$d\theta d\bar{\theta}$ in fact has the dimensions of the inverse of an action. This is 
a sort of {\it dequantization} procedure \cite{wip}.
The inverse procedure of going from (\ref{12-3}) to (\ref{13-1}) is the quantization procedure
and can be somehow achieved by freezing $\theta$ and $\bar{\theta}$ to $0$
%%%
\begin{eqnarray}
\label{13-2}
\displaystyle \langle\Phi^{q}_{\ssc  f};t_{\ssc f}\vert \Phi^q_{\ssc i};t_{\ssc
i}\rangle &=&
 \int {\cal D}^{\prime\prime}\Phi^q{\cal D}\Phi^p
\exp {i\int idt d\theta d{\bar\theta} \,L[\Phi]\, }\nonumber\\
% &\Downarrow
%\mathrm{quantization:}
&\left \Downarrow\rule[3pt]{0pt}{10pt} \right.
&~\theta,{\bar\theta}\rightarrow 0\\
\displaystyle \langle q_{\ssc f};t_{\ssc f}\vert q_{\ssc i};t_{\ssc i}\rangle
&=&
\int {\cal D}^{\prime\prime}q {\cal D}p \exp{\frac{i}{\Delta}\int dt L[\varphi]}. \nonumber
\end{eqnarray}
%%%
The $\Delta$ appearing in (\ref{13-2})  
is a quantity with the dimensions of an action. It will appear because $\theta\bar{\theta}$
has the dimensions of an action and cannot be sent to zero. What can be sent to zero is 
the dimensionless quantity $\theta \bar{\theta}/\Delta $ and this will then 
make $\Delta$ appear in the expression 
(\ref{13-2}). If we identify $\Delta$ with $\hbar$ then the final result 
of the limit $\theta,\bar{\theta}\to 0$ is just the Feynman's QPI (\ref{13-1}).

The procedure outlined in (\ref{13-2}) allows us to obtain also the quantization 
of observables different than the Hamiltonian and representations different than the 
coordinate one. For example, by performing a partial Fourier transform of (\ref{12-1}) 
involving the $q$-variables, we can write the CPI in the $\Phi^p$-representation
in which the operators of multiplication are $(\lambda_q,p,\bar{c}_q,c^p)$:
%%%
\begin{equation}
\displaystyle \label{pip}
\langle \Phi^p_{\ssc f};t_{\ssc f}|\Phi^p_{\ssc i};t_{\ssc i}\rangle=\int
{\cal D}^{\prime\prime} \Phi^p{\cal D}\Phi^q \, \exp\biggl[i\int dt (\lambda_p\dot{p}-q\dot{\lambda}_q
+ic^q\dot{\bar{c}}_q+i\bar{c}_p\dot{c}^p-\widetilde{\cal H})\biggr].
\end{equation} 
%%%
Also in this representation the CPI (\ref{pip}) 
has the peculiar property that its exponential 
weight can be rewritten in terms of the superfields. In fact:
%%%
\begin{equation}
\lambda_p\dot{p}-q\dot{\lambda}_q+ic^q\dot{\bar{c}}_q+i\bar{c}_p\dot{c}^p-\widetilde{\cal H}=
\int id\theta d\bar{\theta} [-\Phi^q\dot{\Phi}^p-H(\Phi)].
\end{equation} 
%%%
Therefore the kernel of propagation becomes:
%%%
\begin{equation}
\label{pip2}
\langle \Phi^p_{\ssc f};t_{\ssc f}|\Phi^p_{\ssc i};t_{\ssc i}\rangle=
\int {\cal D}^{\prime\prime}\Phi^p{\cal D}\Phi^q\,\exp\biggl[i\int idt d\theta d\bar{\theta}[-\Phi^q
\dot{\Phi}^p-H(\Phi)]\biggr].
\end{equation}
%%%
Applying to (\ref{pip2}) the same $\theta,\bar{\theta}\to 0$ procedure used in equation 
(\ref{13-2})  we obtain the following path integral:
%%%
\begin{equation}
\langle p_{\ssc f};t_{\ssc f}|p_{\ssc i};t_{\ssc i}\rangle=\int {\cal D}^{\prime\prime}p
{\cal D}q\; \exp\biggl[\frac{i}{\Delta}\int dt (-q\dot{p}-H)\biggr]
\end{equation}
%%%
which, identifying $\Delta$ with $\hbar$, is just the QPI in the momentum representation.

The quantization procedure outlined in this paper has some ``resemblance" 
with
Geometric Quantization (GQ) \cite{geomquant}. We say so for the following reasons:
\begin{itemize}
\item[1)] GQ postulates a classical Hilbert space which is
called prequantization space \cite{geomquant} and which is basically the KvN space we started from;
\medskip
\item[2)] GQ starts from the Hamiltonian vector field, whose extension is the Lie derivative
of the Hamiltonian flow
$\widetilde{\cal H}$,
and ends with the Schr\"odinger operator. This is exactly what the procedure
(\ref{13-2}) does: we start with the weight $\int idt d\theta d\bar{\theta}L[\Phi]$, which is  
related to the Lie derivative $\widetilde{\cal H}$
of (\ref{8-2}), and we end up via (\ref{13-2}) with the weight 
$\displaystyle \frac{1}{\Delta}\int dt L[\varphi]$ which gives the Schr\"odinger operator;
\medskip
\item[3)] in GQ there is a procedure, called polarization, which reduces the KvN Hilbert space 
with basis $|q,p\rangle$ to the standard quantum Hilbert space. 
We also have a procedure which
brings us from the enlarged states $|\varphi,c\rangle$ first 
to the Fourier transformed ones
$|\Phi^q\rangle=|q,\lambda_p,c^q,\bar{c}_p\rangle$
or $|\Phi^p\rangle=|\lambda_q,p,\bar{c}_q,c^p\rangle$ and then, 
via $\theta,\bar{\theta}\to 0$, to the $|q\rangle$ or the $|p\rangle$ which are two possible
bases for the standard quantum Hilbert space.
\end{itemize}

The beauty of our procedure is that the same two steps (i.e. the Fourier transform and the limit
$\theta,\bar{\theta}\to 0$) which bring the Lie derivative into the Schr\"odinger 
operator (which is point 2) above) also polarize the states (which is point 3) above). 
This is why we find our procedure much more compact 
and even more ``geometrical" than GQ because of the emphasis it puts on the
Grasmannian partners of time. 
As we said before there are only similarities between our procedure and the one of GQ,
because the intrinsic steps and ideology are different. For example in GQ the usual commutator
structure is the one of KvN given by (\ref{comm})
and the operators, not the commutators, are changed in order to get 
the quantum structures. In our procedure instead the commutators (\ref{comm})
are changed from the beginning by the $\theta,\bar{\theta}\to
0$ limit. In fact this limit produces a different kinetic term in the weight of the path integral
and it is this 
kinetic term which generates the commutators. It is most probably this intrinsic difference between GQ and
our procedure which does not give rise in our case to the other serious problem which GQ faces that is the
one of having some observables which cannot be quantized unless one introduces extra structures (metaplectic
stuff, etc.). 

\section{Conclusions}

Beside the resemblance we have noticed with GQ we feel that the most important thing our quantization
procedure has brought to light is that time $t$, or its Grassmannian partners, $\theta$ and $\bar{\theta}$,
play a role in the process of quantization. Via the universal symmetries of the CPI that we
have studied \cite{enniocl}, we know that these $\theta,\bar{\theta}$ can be turned into the normal 
time $t$. So if $\theta$ and $\bar{\theta}$ play a role in quantization also $t$ will do. 
This last issue is under
investigation at the moment.

\section*{Acknowledgments}

The work contained in this paper was presented by one of us (E.G.) at the conference in Vietri
honouring Balachandran on the occasion of his 65th birthday. We thought this paper was particularly
suited for this occasion because the central technical element of this work is the role played by some
particular surface terms. Surface terms have been one of the important issues which Balachandran 
has brought to the attention of the physics community over the last 25 years and, as E.G. said at his 
seminar in Vietri, Balachandran can for sure be qualified as a ``{\it maestro}" of surface terms.
This work was supported by grants from MIUR and INFN of Italy and grants RFBR 03-02-16209
and RFBR NSh-1774.2003.2 of Russia.

\end{document}